\documentclass[prl,aps,
amssymb,
twocolumn,
floatfix,
preprintnumbers,
superscriptaddress,longbibliography,10pt]{revtex4-1}

\usepackage{hyperref}
\usepackage[dvipsnames]{xcolor}
\usepackage{amsmath,amssymb,mathrsfs,amsthm,tikz,shuffle,paralist,accents}
\usepackage{subcaption}
\usepackage{cleveref}
\usepackage{enumitem}

\usepackage{mathtools}
\usepackage{slashed}
\newcommand{\eqnDiag}[1]{ \vcenter{\hbox{ #1}} }

\begin{document}

\title{Feynman Integral Reduction and Landau Singularities}

\author{Federico~Coro}
\email{federico.coro@ugent.be}
\affiliation{Department of Physics and Astronomy, Ghent University, 9000 Ghent, Belgium}

\author{Pavel~P.~Novichkov}
\email{pavel.novichkov@ugent.be}
\affiliation{Department of Physics and Astronomy, Ghent University, 9000 Ghent, Belgium}

\author{Ben~Page}
\email{ben.page@ugent.be}
\affiliation{Department of Physics and Astronomy, Ghent University, 9000 Ghent, Belgium}

\author{Qian~Song}
\email{qian.song@ugent.be}
\affiliation{Department of Physics and Astronomy, Ghent University, 9000 Ghent, Belgium}

\begin{abstract}

  We propose that Feynman integral reduction is controlled by solutions of
  the Landau equations.
  We study integral relations with prescribed propagator powers using syzygy
  methods and discuss how syzygies can be expressed as a sum over
  components of the Landau singularity locus.
  This leads to a determinantal approach to solving the syzygy problem, giving rise to
  highly compact and physically transparent solutions.
  We demonstrate the
  method in applications to planar two-loop five-point integrals relevant for the $pp
  \rightarrow t\overline{t}H$ process.
  Our results suggest an efficient method of Feynman integral reduction and
  provide a novel physical perspective on the problem.

\end{abstract}

\maketitle


\section{Introduction}

Perturbative calculations of scattering amplitudes are essential for precise
understanding of fundamental physics at particle collider and gravitational-wave
experiments.
The defining difficulty of such calculations is both the
large number and complexity of the Feynman integrals that must be computed.
In modern approaches, this issue is tackled by exploiting the linear
relations that the integrals satisfy~\cite{Chetyrkin:1981qh}.
This both reduces the computational burden to that of a smaller set of ``master
integrals'' as well as facilitates the calculation of the master integrals
themselves via the differential equations
method~\cite{Kotikov:1990kg,Remiddi:1997ny,Gehrmann:1999as,Henn:2013pwa}.
While algorithms for integral reduction exist~\cite{Laporta:2000dsw} and public
implementations are available~\cite{
Maierhofer:2017gsa,Klappert:2020nbg,
Smirnov:2008iw,Smirnov:2025prc,
Peraro:2019svx,
Guan:2024byi,
Lee:2012cn,Lee:2013mka,
Wu:2023upw,Wu:2025aeg
}, the large demands for precision placed by
experimental uncertainties have driven the development of novel reduction
methods such as syzygy methods~\cite{Gluza:2010ws,Larsen:2015ped,Ita:2015tya},
finite-field reconstruction~\cite{vonManteuffel:2014ixa,Peraro:2016wsq} and
twisted
cohomology approaches~\cite{Mizera:2017rqa, Mastrolia:2018uzb, Mizera:2019vvs,
Frellesvig:2020qot, Caron-Huot:2021xqj, Caron-Huot:2021iev}.

In this letter we argue that Feynman integral
reduction is controlled by the Landau equations~\cite{Landau:1959fi}.
The Landau equations have recently been under intense
study~\cite{Helmer:2025yuf,Helmer:2025ljj,Mizera:2021icv,Fevola:2023kaw,Fevola:2023fzn,Correia:2025yao,Helmer:2024wax,Dlapa:2023cvx}
as singularities of scattering amplitudes are described by solutions that constrain external kinematics.
In this work we discuss how solutions of the Landau equations that do not
constrain external kinematics organize the set of relations
between Feynman integrals.
We uncover this physical correspondence by considering the set of integral
relations with prescribed propagator powers.
It is well-known that these relations are controlled by a so-called syzygy
equation~\cite{Gluza:2010ws}.
Despite intense study over the past 15
years~\cite{Bern:2017gdk,Abreu:2017xsl,Abreu:2017xsl,Zhang:2016kfo,Bohm:2017qme,Bosma:2017hrk,Larsen:2015ped,Schabinger:2011dz,Ita:2015tya,Abreu:2023bdp,Agarwal:2020dye,Gluza:2010ws}
and impressive recent computational progress~\cite{Bendle:2019csk,Wu:2023upw,Wu:2025aeg},
both the physical meaning of the syzygy equation and general solutions
have remained elusive.
In this work, we tackle both questions by connecting the syzygy problem to
the Landau equations.

While a link between syzygies and the Landau equations has been noticed
before~\cite{Bern:2017gdk}, we employ two advances to make this connection
constructive.
First, we exploit the recent observation~\cite{Page:2025gso} that the
large-$\epsilon$ limit~\cite{Mizera:2019vvs} distinguishes so-called ``critical
syzygies'' for integral reduction purposes.
Second, we use advanced technologies from commutative
algebra~\cite{AtiyahMacdonald1969,eisenbud1995commutative}. Together, this
allows us to write the set of critical syzygies as a sum over
components of the solutions of the Landau equations.
Importantly, this leads to an approach to construct a highly-compact collection of
analytic syzygies.
We illustrate our approach by applying it to a cutting-edge example of a
two-loop pentabox family for the frontier $pp \rightarrow t\overline{t}H$
process. We discuss a double box case in detail and present the remaining set of
five-point syzygy solutions in supplemental material.
Altogether, our results highlight how Feynman integral reduction is controlled
by infrared singularities and provide an efficient method of syzygy construction
for phenomenological application.

\section{Singularities and Integral Relations}

In this work, we study relations between Feynman integrals with prescribed
propagator powers. It is well known~\cite{Gluza:2010ws} that these relations are
easily constructed once a ``syzygy equation'' is solved.
In the Baikov representation~\cite{Baikov:1996rk},
this equation can be formulated as~\cite{Larsen:2015ped,Page:2025gso}
\begin{equation}
   0 = a_0 B + \sum_{i \in \text{ISPs}(\Gamma)}\!\!\!\! a_i \partial_i B + \!\sum_{e \in \text{props}(\Gamma)} \!\!\!\!\!\! \left( \tilde{a}_ez_e B \!+\! \overline{a}_e z_e \partial_e B \right),
   \label{eq:MasterSyzygy}
\end{equation}
where $\Gamma$ is the graph associated to the Feynman integral, $\text{ISPs}(\Gamma)$ labels the irreducible
scalar product variables and $\text{props}(\Gamma)$ labels the propagator
variables $z_e$.
In \cref{eq:MasterSyzygy}, $B$ is the Baikov polynomial which, in two-loop,
five-point applications, can be normalized and written as
\begin{equation}
  B = \mu_{11} \mu_{22} - \mu_{12}^2, \qquad
  \mu_{ij} = \frac{
    G\left(
      \begin{smallmatrix}
        \!\overline{\ell}_i\! & \!p_1 \! & \!p_2 \! & \! p_3 \! & \! p_4 \! \\
        \!\overline{\ell}_j\! & \!p_1 \! & \!p_2 \! & \! p_3 \! & \! p_4 \!
     \end{smallmatrix}
     \right)
             }{G(p_1 p_2 p_3 p_4)},
  \label{eq:BaikovDef}
\end{equation}
where $\overline{\ell}_i$ are $D$-dimensional loop momenta, $p_k$ are external
momenta, $G(\begin{smallmatrix}V \\ W\end{smallmatrix})$ is the determinant of
the matrix of dot products of vectors in $V$ and $W$ and $G(V) =
G(\begin{smallmatrix}V \\ V\end{smallmatrix})$ .

Despite being a linear equation, solving \cref{eq:MasterSyzygy} is highly non-trivial
since integral reduction demands solutions $a_0$, $a_i$, $\tilde{a}_e$ and
$\overline{a}_e$ that are elements of $R$, the
polynomial ring in Baikov variables.
In ref.~\cite{Page:2025gso}, it was observed that integral relations
induced by solutions to \cref{eq:MasterSyzygy} are governed by the $a_0$
term. This led to the definition of the set of critical syzygy solutions,
$\text{CSyz}(\Gamma)$, that control this feature at the level of
\cref{eq:MasterSyzygy}.
In the following, we develop a novel approach to finding critical syzygy
solutions, by highlighting a connection to the Landau equations.

To begin, we consider the maximal cut 
Landau locus in Baikov parameterization~\cite{Correia:2025yao,Caron-Huot:2024brh}. This is the variety
$U_{\text{Landau}}^\Gamma = V(J^\Gamma_{\text{Landau}})$ associated to the polynomial ideal
\begin{equation}
J^\Gamma_{\text{Landau}} = \langle B \rangle + \sum_{i \in \text{ISPs}(\Gamma)} \langle  \partial_i B \rangle  +  J_{\text{cut}}^\Gamma,
\label{eq:LandauDefinition}
\end{equation}
where $+$ denotes the sum of polynomial ideals and $J_{\text{cut}}^\Gamma = \sum_{e \in \text{props}(\Gamma)} \langle z_e
\rangle$. We consider $U_{\text{Landau}}^\Gamma$ for fixed
external kinematics and hence as a variety in Baikov variables alone.
Physically, $U_{\text{Landau}}^\Gamma$ describes integration space regions
that potentially give rise to divergences.
For two-loop, five-point examples, we find that $U_{\text{Landau}}^\Gamma$ admits a
decomposition into subvarieties as
\begin{equation}
U^\Gamma_{\text{Landau}} = V(J_{\text{4d}} + J_{\text{cut}}^\Gamma) \cup \left[ \bigcup_{i=1}^{n_{\text{IR}}} U^\Gamma_i \right],
\label{eq:LandauDecomposition}
\end{equation}
where $J_{\text{4d}} = \langle  \mu_{11}, \mu_{22}, \mu_{12} \rangle$
and each $U_i^\Gamma$ is an irreducible component
of $U^\Gamma_{\text{Landau}}$ that is
not contained in $U_0^\Gamma = V(J_{\text{4d}} + J_{\text{cut}}^\Gamma)$.
We observe that, physically, the $U_{i>0}$ are associated to soft and collinear
configurations.

Our main result is to write the critical syzygy module as a sum over
the varieties in \cref{eq:LandauDecomposition} as
\begin{equation}
  \text{CSyz}(\Gamma) = \sum_{i=0}^{n_{\text{IR}}} \text{FittSyz}(U^\Gamma_i),
  \label{eq:CriticalSyzygyDecomposition}
\end{equation}
which we dub a ``Landau decomposition'' of $\text{CSyz}(\Gamma)$.
In the following, we define $\text{FittSyz}(U_i^\Gamma)$ by giving an
explicit procedure to construct a generating set.
To this end, we recall that the module of critical syzygies,
$\text{CSyz}(\Gamma)$, is defined such that it is isomorphic to an
ideal~\cite{Page:2025gso} as
\begin{gather}
    \text{CSyz}(\Gamma) \simeq A_0^\Gamma/J_{\text{cut}}^\Gamma, \qquad \qquad A_0^\Gamma = J_{\text{syz}}^\Gamma : \langle  B \rangle,
  \label{eq:CSyzIsomorphism}
    \\
  J_{\text{syz}}^\Gamma = \sum_{i \in \text{ISPs}(\Gamma)} \langle \partial_i B \rangle + \sum_{e \in \text{props}(\Gamma)} \langle  z_e B, z_e \partial_e B \rangle,
  \label{eq:JSyzDef}
\end{gather}
where $A_0^\Gamma/J_{\text{cut}}^\Gamma$ is the image of $A_0^\Gamma$ in the
quotient ring  $R/J_{\text{cut}}^\Gamma$ and $J^{\Gamma}_{\text{syz}}:\langle B \rangle$ is an ideal quotient.
Geometrically, if this quotient is saturated, i.e. $J_{\text{syz}}^\Gamma : \langle B
\rangle = J_{\text{syz}}^\Gamma : \langle B^2 \rangle$, then it can be
understood as removing components where $B=0$ from $V(J_{\text{syz}}^\Gamma)$~\cite{Page:2025gso}.
We will make use of this ``saturation hypothesis'' throughout this work.
Here, we interpret the ideal quotient as removing Landau singularities from
$V(J_{\text{syz}}^\Gamma)$ and
\cref{eq:CriticalSyzygyDecomposition} as an algebraic shadow of this
geometric intuition.
That is, one can remove Landau locus components one by one in a way that
automatically yields critical syzygies.

To set up \cref{eq:CriticalSyzygyDecomposition} and introduce some technology.
The main observation is that we can decompose $A_0^\Gamma$ as
\begin{align}
  A_0^\Gamma  = &\sum_{i=0}^{n_{\text{IR}}} \text{Fitt}_0 \! \left( M_i \right) + J_{\text{cut}}^\Gamma,
  \label{eq:A0Decomposition}
  \\
  M_0 = J_{\text{4d}} / J_{\text{syz}}^\Gamma, &\qquad M_i = \left(J_{\text{syz}}^{\Gamma, [U^\Gamma_i]} \!\cap \!J_{\text{4d}} \right) / J_{\text{syz}}^\Gamma,
\end{align}
where the $M_i$ are viewed as $R$-modules and the $i = 0$ term in both \cref{eq:A0Decomposition} and
\cref{eq:CriticalSyzygyDecomposition} can be dropped if $n_{\text{IR}} > 0$.
This result, and thus \cref{eq:CriticalSyzygyDecomposition}, can be shown
to hold under the saturation hypothesis and
\begin{equation}
V\left(J_{\text{syz}}^\Gamma : \sqrt{J_{\text{syz}}^\Gamma}\right) \cap U^\Gamma_{\text{Landau}} \subseteq \bigcup_{i=1}^{n_\text{IR}} U^\Gamma_i,
\label{eq:RadicalityCondition}
\end{equation}
where $\sqrt{J}$ denotes the radical of $J$.
$V(J : \sqrt{J})$ is known as the non-reduced locus of $J$ and describes the
locus where $J$ is not a radical ideal.
The proof is technical and we leave it to
supplemental material.

There are two main pieces of mathematical technology in \cref{eq:A0Decomposition}. The first is
$\text{Fitt}_0(M_i)$, the zeroth Fitting ideal of the module
$M_i$~\cite{eisenbud1995commutative}. Denoting a generating set of the
numerator ideal of $M_i$ as $\{v_{i,1}, \ldots, v_{i,n}\}$,
$\text{Fitt}_0(M_i)$ is generated by the $n \times n$ minors of a
presentation matrix $\mathcal{M}_i$ of $M_i$:
a matrix whose rows generate the module of syzygies of a generating set of $M_i$.
Geometrically, for two ideals $J$, $K$ such that $J \subset K$, the zeroth Fitting ideal
satisfies
\begin{equation}
  V(\text{Fitt}_0[K/J]) = V(J : K),
  \label{eq:FittingIdealGeometry}
\end{equation}
and is thus intimately related to the ideal quotient.
The second technology is $J^{[U]}$, the ``pre-localization'' of an ideal $J$ at a variety
$U$, which can be built from a primary decomposition of $J$.
Denoting the primary components of $J$ as $\mathcal{Q}_{j}$, we define the
pre-localization as those components whose associated variety intersects $U$,
i.e.
\begin{equation}
   J^{[U]} = \bigcap_{j : V(\mathcal{Q}_j) \cap U \ne \emptyset } \mathcal{Q}_{j}.
\end{equation}
Thus, 
each term in \cref{eq:A0Decomposition} excises $U_0^\Gamma \cup U_i^\Gamma$
from $V(J_{\text{syz}}^\Gamma)$.

In practice, we can view \cref{eq:A0Decomposition} as a determinantal
construction of $A_0^\Gamma$.
Moreover, via judicious, block construction of the $\mathcal{M}_i$, we can also use this to build
$\text{CSyz}(\Gamma)$.
To this end,
writing the generators of $J_{\text{syz}}^\Gamma$ in \cref{eq:JSyzDef} as
$\{b_1, \ldots, b_m\}$, we construct a polynomial matrix $\mathcal{C}_{i}$ such
that $b_j = (\mathcal{C}_{i})_{jk} v_{i,k}$.
Next, we construct a matrix $\mathcal{R}_{i}$ whose rows generate the module
of syzygies of 
$\{v_{i, 1}, \ldots, v_{i,n}\}$.
It can be shown that the augmented matrix $\mathcal{M}_i = \left(\begin{smallmatrix}
\mathcal{C}_{i} \\ \mathcal{R}_{i} \end{smallmatrix} \right)$ is a presentation matrix
of $M_i$.
To use this to construct syzygies, we exploit that, under the saturation hypothesis,  there
exists a $\vec{c}_{i}$ such that $B = \vec{c}_{i} \cdot \vec{v}_i$.
We can combine this vector and $\mathcal{M}_i$ into the matrix
\begin{equation}
  \mathcal{F}_i =
  \left(
  \begin{array}{cc}
    B & \vec{c}_i^T \\
    \vec{b} & \mathcal{C}_{i} \\
    \vec{0} & \mathcal{R}_{i}
  \end{array}
  \right),
  \label{eq:FittingSyzygyMatrix}
\end{equation}
which satisfies $\mathcal{F}_i\left( \begin{smallmatrix} -1 \\
  \vec{v}_i \end{smallmatrix} \right)  = 0$ and so maximal minors of
$\mathcal{F}_i$ vanish.
Therefore, Laplace expansion along the leftmost column yields solutions to
\cref{eq:MasterSyzygy}.
We define $\text{FittSyz}(U_i^\Gamma)$ as the module implicitly generated by Laplace
expansions of the maximal minors of $\mathcal{F}_i$ that contain the first row.
By construction, the coefficients of $B$ generate $\text{Fitt}_0(M_i)$ and we
conclude that the decomposition of \cref{eq:CriticalSyzygyDecomposition} holds.

In summary, to construct the set of critical syzygies for a topology $\Gamma$,
we first construct an irreducible decomposition of the maximal-cut Landau locus.
Then, for each component, we construct the matrix $\mathcal{F}_i$ by computing
the set of polynomial matrices $c_i$, $\mathcal{C}_i$ and $\mathcal{R}_i$.
Laplace expansion of each $\mathcal{F}_i$ gives syzygy solutions which, when
taken together, generate the syzygy module $\text{CSyz}(\Gamma)$.

\section{\bf Notation and Variable Choice}

In the rest of this letter, we demonstrate how \cref{eq:FittingSyzygyMatrix} can
be used to efficiently build syzygy solutions.
For concreteness, we study applications to cases arising in the two-loop
contributions to amplitudes for $pp \rightarrow t\overline{t}H$.
Here, the external momenta satisfy
\begin{equation}
  \sum_{i=1}^5 p_i = 0, \qquad p_{1,3}^2 = m_t^2, \quad p_2^2 = m_H^2, \quad p_{4,5}^2 = 0.
\end{equation}
We route the loop momenta as in \cref{fig:DB} and, writing $q_k = \sum_{j=1}^k
p_j$, $q_0 = 0$,
we define Baikov variables as
\begin{equation}
  z_{i,j} = (\overline{\ell}_i - q_j)^2 - m_{i,j}^2, \qquad
  \tilde{z} = (\overline{\ell}_1 - \overline{\ell}_2)^2,
  \label{eq:BaikovFromLoopMomenta}
\end{equation}
where $m_{i,j}$ is the mass of the propagator
associated to $z_{i,j}$.
We abbreviate $\partial_{a,j} = \frac{\partial}{\partial z_{a,j}}$ and
$\partial_{\tilde{z}} = \frac{\partial}{\partial \tilde{z}}$.

For practical computations, as well as ease of physical understanding, we make the
observation that, for fixed external kinematics,
\cref{eq:BaikovFromLoopMomenta} defines an isomorphism between
the polynomial ring in Baikov variables and a
polynomial ring in momentum variables. Specifically,
\begin{equation}
  R \cong \mathbb{C}[\ell_1^\mu, \ell_2^\mu, \mu_{11}, \mu_{22}, \mu_{12}],
  \label{eq:RingIsomorphism}
\end{equation}
where $\ell_i$ are the four-dimensional projections of the $\overline{\ell}_i$
and $\mu_{ij} = \overline{\ell}_i \cdot \overline{\ell}_j - \ell_i \cdot
\ell_j$.
The isomorphism in \cref{eq:RingIsomorphism} follows as both the forward and the
inverse map are polynomial, with
inverse map given by the well-known
\begin{equation}
  \ell_{i} = \frac{1}{2} \sum_{j} \left( z_{i,0} - z_{i,j}  + q_j^2 - m_{i,j}^2 + m_{i,0}^2 \right) v_j,
\end{equation}
where the $v_i$ are defined by $v_i \cdot q_j = \delta_{ij}$, alongside \cref{eq:BaikovDef}.

The Baikov polynomial and its derivatives can easily be expressed in
loop-momentum variables via \cref{eq:BaikovDef} and 
\begin{align}
  \partial_{\tilde{z}} B &= \mu_{12},
  \label{eq:dZmu12}
  \\
  \begin{split}
  \partial_{a,j} B &= \delta_{j,0}(\mu_{\overline{a}\overline{a}} - \mu_{12}) + w_j \cdot \Lambda_{a},
  \\
  \Lambda_{a}^\nu &= \ell_{a}^\nu \mu_{\overline{a}\overline{a}} - \ell_{\overline{a}}^\nu \mu_{12},
  \end{split}
  \label{eq:BaikovDerivativesMomenta}
\end{align}
where $\overline{a} = 3-a$ and $w_i^\mu = -2 \partial_{a,i} \ell_a^\mu$.
Moreover, we make use of the Sudakov parameterization,
\begin{equation}
  \ell_i = x_{i} p + \beta_{i} \eta + \ell_i^{\perp}
  \label{eq:SudakovParameterization}
\end{equation}
where $p$ is the relevant massless momentum, $\eta$ is a massless reference
vector and $\ell_i^\perp$ is the projection of $\ell_i$ into the space
transverse to $p$ and $\eta$.

\section{A Five-Point Double Box Example}
\begin{figure}[t]
  \begin{subfigure}{0.15\textwidth}
    $\eqnDiag{\includegraphics[scale=0.45]{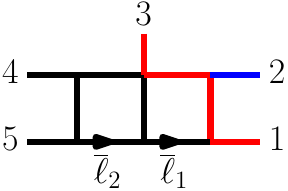}}$
  \caption{$\Gamma_1$}
  \label{fig:DB}
  \end{subfigure}
  \begin{subfigure}{0.15\textwidth}
    $\eqnDiag{\includegraphics[scale=0.45]{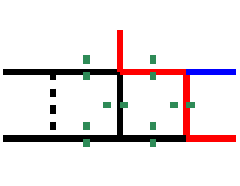}}$
  \caption{$U_{1}^{\Gamma_1}$}
  \label{fig:DBU1}
  \end{subfigure}
  \begin{subfigure}{0.15\textwidth}
    $\eqnDiag{\includegraphics[scale=0.45]{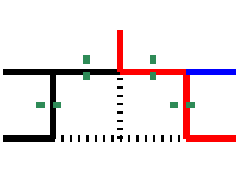}}$
  \caption{$U_{2}^{\Gamma_1}$}
  \label{fig:DBU2}
  \end{subfigure}

  \caption{Double box topology from $pp \rightarrow
    t\overline{t}H$ and its leading Landau singularities. Black lines
    are massless particles; red, the top quark; blue, the Higgs boson; dashed,
    soft particles; dotted, collinear particles.
  }
  \label{fig:FivePointExamples}
\end{figure}

In this section, we apply \cref{eq:CriticalSyzygyDecomposition} in the
case of the double box of \cref{fig:DB}. All explicit computations are performed with
\texttt{Singular}~\cite{DGPS}.
To begin, we check the validity of the Landau decomposition, by computing the
ingredients of \cref{eq:RadicalityCondition}.
We first compute the irreducible decomposition of the maximal cut Landau locus, finding
\begin{equation}
  U^{\Gamma_1}_{\text{Landau}} = U_{0}^{\Gamma_1} \cup U_{1}^{\Gamma_1} \cup U_{2}^{\Gamma_1},
\end{equation}
where $U_1^{\Gamma_1}$ and $U_2^{\Gamma_2}$ correspond to configurations of loop-momenta that both satisfy
the maximal cut, as well as some internal soft/collinear configuration. We
depict them graphically in~\cref{fig:DBU1,fig:DBU2}.

Next, to compute the non-reduced locus of $J_{\text{syz}}^{\Gamma_1}$, we
calculate the radical of $J_{\text{syz}}^{\Gamma_1}$. Our approach is to express
it as the intersection
of the minimal associated primes~\cite{AtiyahMacdonald1969} of $R/J_{\text{syz}}^{\Gamma_1}$, which we
write as $\text{mAss}\left(J_{\text{syz}}^{\Gamma_1}\right)$. As minimal
primes correspond to irreducible components of
$V(J_{\text{syz}}^\Gamma)$, and generators of $J_{\text{syz}}^\Gamma$
factorize, it can be argued that
\begin{equation}
  \text{mAss} \!\left( J_{\text{syz}}^{\Gamma} \right)
  \!=\! \text{mAss}\! \left( A_0^\Gamma \right) \! \cup
\text{min}_{\subseteq} \! \left[
      \bigcup_{\Gamma' \subseteq \Gamma} \! \text{mAss}\!\left( J_{\text{Landau}}^{\Gamma'} \right)
    \right] \! , 
    \label{eq:MinAssocFormula}
\end{equation}
where the minimum is with respect to inclusion.
For the two-loop examples we have studied, \cref{eq:MinAssocFormula} can be efficiently
implemented for numerical external kinematics.

\begin{figure}
  \begin{subfigure}{0.15\textwidth}
      $\eqnDiag{\includegraphics[scale=0.5]{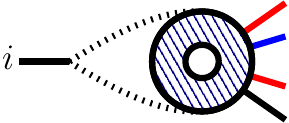}}$
  \caption{$p_i$ collinear}
  \label{fig:ExternalCol}
  \end{subfigure}
  \begin{subfigure}{0.15\textwidth}
      $\eqnDiag{\includegraphics[scale=0.5]{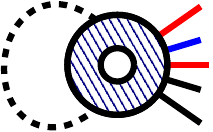}}$
  \caption{soft}
  \label{fig:SingleSoft}
  \end{subfigure}
  \begin{subfigure}{0.15\textwidth}
    $\eqnDiag{\includegraphics[scale=0.5]{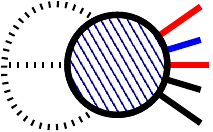}}$
  \caption{$\overline{\ell}_1/\overline{\ell}_2$ collinear}
  \label{fig:InternalSingularity}
  \end{subfigure}
  \caption{Diagrams of Landau
    singularities arising in the computation of $\sqrt{J_{\text{syz}}^\Gamma}$. Notation matches \cref{fig:FivePointExamples}.}
  \label{fig:FundamentalSingularities}
\end{figure}

For the double box example, we find the result
\begin{equation}
  \sqrt{\!J_{\text{syz}}^{\Gamma_1}} \!\mkern-1mu = \!\mkern-1mu A_0^{\Gamma_1} \mkern-1mu \cap \mkern-1mu J_{\text{4d}} \cap J_{\text{e-col}}^{\ell_2, p_5} \cap J_{\text{e-col}}^{\ell_2 \mkern-1mu-\mkern-1mu q_4, p_4} \cap J_{\text{soft}}^{\ell_1 \mkern-1mu-\mkern-1mu \ell_2} \cap J_{\text{soft}}^{\ell_1} \cap J_{\text{i-col}}^{\ell_1\mkern-1mu, \ell_2}\mkern-2mu.
  \label{eq:DoubleBoxRadical}
\end{equation}
The novel ideals in \cref{eq:DoubleBoxRadical} correspond to uncut soft or
collinear singularities, which we depict in
\cref{fig:FundamentalSingularities}.
Two ideals describe configurations where an internal momentum is collinear to an
external momentum, as in \cref{fig:ExternalCol}, two ideals describe a soft
internal momentum, as in \cref{fig:SingleSoft} and a final ideal describes three
collinear internal momenta, as in \cref{fig:InternalSingularity}.
Explicitly, they are given by
\begin{align}
  J_{\text{e-col}}^{\ell_i - q_j, p_k} &= \langle \mu_{12}, \mu_{ii}, (\ell_i - q_j)^{\perp \nu}, (\ell_i - q_j) \cdot p_k \rangle, \\
  J_{\text{soft}}^{\ell_1 } &= \langle \ell_1^\nu, \mu_{11}, \mu_{12} \rangle,\\
  J_{\text{soft}}^{\ell_1 - \ell_2} &= \langle \ell_1^\nu - \ell_2^\nu, \mu_{11} - \mu_{12}, \mu_{22} - \mu_{12} \rangle,\\
  J^{\ell_1,\ell_2}_{\text{i-col}} &= \langle \ell_1^\mu \ell_2^\nu - \ell_1^\nu \ell_2^\mu,
  z_{1,0},
  z_{2,0},
  \tilde{z}
  \rangle,
\end{align}
where we employ the parameterization of \cref{eq:SudakovParameterization} for $p =
p_k$.
With this data, we compute the non-reduced locus of $J_{\text{syz}}^{\Gamma_1}$
and find that its intersection with the maximal cut is
empty, implying that \cref{eq:RadicalityCondition} is satisfied.

These ideals allow us to easily compute the necessary pre-localizations and
their associated $\mathcal{C}$ matrices.
As the non-reduced locus does not intersect the maximal cut, all
pre-localizations are simply intersections of the ideals in
\cref{eq:DoubleBoxRadical}.
For $U_{1}^{\Gamma_1}$, it is physically
clear that it intersects only the $p_4/p_5$ collinear loci and the 4d locus, giving
\begin{align}
  \begin{split}
  J_{\text{syz}}^{\Gamma_1, [U_{1}^{\Gamma_1}]} &= J_{\text{4d}} \cap J_{\text{e-col}}^{\ell_2, p_5} \cap J_{\text{e-col}}^{\ell_2 - q_4, p_4}
  \\
  &=\langle \mu_{12}, \mu_{22}, \mu_{11}  \ell_2^{\perp \nu}, z_{2,0} z_{2,3} \mu_{11},  z_{2,4} \mu_{11} \rangle,
  \end{split}
  \label{eq:ExternalSoft}
\end{align}
where we employ the parameterization of \cref{eq:SudakovParameterization} with $p = p_5, \eta = p_4$.
To construct $\mathcal{C}$, we note that $J_{\text{syz}}^{\Gamma_1,
[U_{1}^{\Gamma_1}]}$ contains $J_{\text{4d}}$.
We express the generators of
$J_{\text{syz}}^{\Gamma_1}$ in terms of those of \cref{eq:ExternalSoft} by
inspection, using \cref{eq:dZmu12} and
\begin{align}
  \partial_{1,j} B &\!= (\partial_{1,j} \mu_{11}) \mu_{22} - 2 (\partial_{1,j} \mu_{12}) \mu_{12},
  \\
  z_{2,4} \partial_{2,4} B &\!= (\partial_{2,4} \mu_{22}) z_{2,4} \mu_{11} - 2 z_{2,4} (\partial_{2,4} \mu_{12}) \mu_{12},
\\
  \partial_{2, j} B &\!= \left( w_j \!\cdot\! \ell_2^\perp \! \right) \! \mu_{11} \!-\! 2 (\partial_{2,j} \mu_{12}) \mu_{12},
\label{eq:SoftOtherDerivatives}
\\
  z_{2,j} \partial_{2,j} B
  &\!=\! z_{2,j} \!\! \left[ \! \frac{z_{2,(3-j)} \!\!-\! z_{2,4}}{2 p_4 \cdot p_5} \!+\! w_j \!\cdot\! \ell_2^\perp \!\right]\!\! \mu_{11} \!-\! z_{2,j} \partial_{2,j} \mu_{12}^2,
  \label{eq:softJetLineDerivatives}
\end{align}
where \cref{eq:SoftOtherDerivatives} holds for $j=1,2$; \cref{eq:softJetLineDerivatives} for $j = 0, 3$.

The pre-localization at $U_{2}^{\Gamma_1}$ can be handled analogously.
Physically, the maximal-cut $\ell_1/\ell_2$ collinear singularity intersects
with the four-dimensional locus, as well as uncut soft and collinear loci,
implying
\begin{align}
  \begin{split}
  &J_{\text{syz}}^{\Gamma_1, [U_{2}^{\Gamma_1}]} = J_{\text{4d}} \cap J_{\text{soft}}^{\ell_1 - \ell_2} \cap J_{\text{i-col}}^{\ell_1, \ell_2},
  \\
  &\phantom{J_{\text{syz}}^{\Gamma_1, [U_{2}^{\Gamma_1}]}} = \langle \Lambda_1^\nu, \chi_{1}^{(a)}, \chi_{1}^{(b)}, B \rangle + (1 \leftrightarrow 2),
  \\
\chi_i^{(a)} &= z_{i,0}(\mu_{\overline{i}\overline{i}} - \mu_{12}), \qquad \chi_i^{(b)} = z_{i,0}(\mu_{ii} - \mu_{\overline{i}\overline{i}}).
  \end{split}
  \label{eq:InternalSoftCol}
\end{align}
The relations required to express the $\mathcal{C}$ matrix are again easy to
determine by inspection. They are given by 
\begin{align}
  z_{i,0} \partial_{i,0} B &= z_{i,0} w_0 \cdot \Lambda_i + \chi_i^{(a)},
  \label{eq:dzi0LambdaMaRelation}
  \\
  \tilde{z} \partial_{\tilde{z}} B &= 2 B + \ell_1 \cdot \Lambda_1 + \ell_2 \cdot \Lambda_2 - \chi_1^{(a)} - \chi_2^{(a)},
\end{align}
alongside \cref{eq:BaikovDerivativesMomenta} for $j>0$.
Given the generating sets in \cref{eq:ExternalSoft,eq:InternalSoftCol}, the
syzygy matrices $\mathcal{R}$ can easily be computed with Groebner basis techniques.

\section{$pp \rightarrow t\overline{t}H$ Pentabox Study}

\begin{figure}
  \begin{subfigure}{0.085\textwidth}
    $\eqnDiag{\includegraphics[scale=0.35]{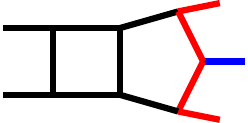}}$
  \caption{$\Gamma_2$}
  \label{fig:PB}
  \end{subfigure}
  \begin{subfigure}{0.085\textwidth}
    $\eqnDiag{\includegraphics[scale=0.35]{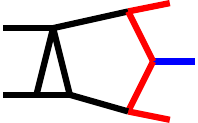}}$
  \caption{$\Gamma_3$}
  \label{fig:PT}
  \end{subfigure}
  \begin{subfigure}{0.085\textwidth}
    $\eqnDiag{\includegraphics[scale=0.35]{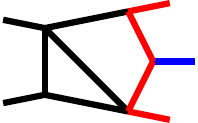}}$
  \caption{$\Gamma_4$}
  \label{fig:BT}
  \end{subfigure}
  \begin{subfigure}{0.085\textwidth}
    $\eqnDiag{\includegraphics[scale=0.35]{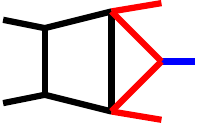}}$
  \caption{$\Gamma_5$}
  \label{fig:BT2}
  \end{subfigure}
  \begin{subfigure}{0.07\textwidth}
    $\eqnDiag{\includegraphics[scale=0.35]{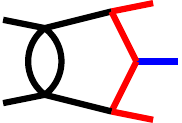}}$
  \caption{$\Gamma_6$}
  \label{fig:PBub}
  \end{subfigure}

  \caption{Remaining five-point topologies contributing to the leading-color,
    light-quark-loop contributions to $pp \rightarrow
    t\overline{t}H$ pentabox topology.}
  \label{fig:ExtraExamples}
\end{figure}

We have applied the Landau decomposition strategy to the set of five-point
diagrams within the pentabox topology of \cref{fig:PB}, which we depict in
\cref{fig:ExtraExamples}.
In all cases, similar to the double box example, we confirm that the non-reduced
locus of $J_{\text{syz}}^\Gamma$ does not intersect the maximal cut, therefore
implying \cref{eq:RadicalityCondition}. As such, we could construct all Landau
decompositions with the strategy described for the double box.
Moreover, we explicitly confirm the equality of \cref{eq:A0Decomposition} computationally.

We find that the universality of infrared singularities is reflected in our
construction: all (singular) minimal primes that we encounter are either $J_{\text{4d}}$ or
are captured by the graphs in
\cref{fig:FundamentalSingularities}.
Moreover, we are able to recycle ingredients between topologies.
Firstly, the ideal in \cref{eq:ExternalSoft} arises as a pre-localization on the soft
singularity for an exchange between the two massless legs also in the pentabox,
$\Gamma_2$, and the box-triangle, $\Gamma_5$.
Secondly, for each of $\Gamma_4$, $\Gamma_5$ and $\Gamma_6$, the singularity
where the central line is soft results in the same pre-localization.
In both cases, one can easily reuse the building blocks of the Fitting construction.
In total, only 7 non-trivially distinct constructions arise.
For the interested reader, we present results of the calculations of the
$\mathcal{C}$ and $\mathcal{R}$ matrices for all graphs in
\cref{fig:ExtraExamples} in supplemental material.


\section{\bf Discussion and Outlook}

In this work, we have studied Feynman integral reduction with syzygy methods. We applied advanced methods from commutative algebra to
construct analytic generating sets of syzygies that are essential for efficient
computation of Feynman integrals for collider physics.
Our formulae are highly compact, and physically transparent: they are organized
by the Landau locus.
Strikingly, this highlights how Feynman integral reduction is controlled by
infrared singularities.
Interestingly, we have observed that, once syzygies associated to a
Landau locus component have been constructed, they can be applied to other diagrams.
This suggests a universality of the syzygy solutions that we have presented.

Given our results, an important next step is to understand how to construct a
minimal set of total derivatives suitable for integrand-level
reduction~\cite{Ita:2015tya,Abreu:2017xsl}.
Indeed, the compact nature of our results hints at the tantalizing possibility
of constructing stable, floating-point reduction methods in which there has been
renewed
interest~\cite{Abreu:2017xsl,Abreu:2020xvt,Pozzorini:2022ohr,Bevilacqua:2025xms}.
Altogether, we expect our approach to be of crucial importance for future
calculations of LHC scattering amplitudes, enabling progress both beyond
$2\rightarrow 3$ kinematics and at higher loop orders.

\vspace{.14cm}
\noindent {\bf Acknowledgments}
\vspace{.1cm}
We thank Harald Ita, Andrzej Pokraka and Andrew McLeod for insightful discussions.
We thank Harald Ita and Vasily Sotnikov for comments on the draft.
The work of Federico Coro, Pavel Novichkov and Qian Song was supported by the
European Research Council (ERC) under the European Union’s Horizon Europe
research and innovation program grant agreement 101078449 (ERC Starting Grant
MultiScaleAmp).
Views and opinions
expressed are however those of the authors only and do not necessarily reflect
those of the European Union or the European Research Council Executive Agency.
Neither the European Union nor the granting authority can be held responsible
for them.

\appendix

\section{Proof of Fitting Decomposition of $A_0^\Gamma$}

Our proof of \cref{eq:A0Decomposition} makes use of the fundamental concept in
commutative algebra of localization.
We refer the interested reader to \cite[Chapter 2]{eisenbud1995commutative} and
\cite[Chapter 3]{AtiyahMacdonald1969} for introductory details.
The main idea that we use is that ideal containment, and therefore
equality, is a local property. That is, for ideals $J$ and $K$,
\begin{equation}
  \!\!J \mkern-1mu=\mkern-1mu K \,\,\, \Leftrightarrow \,\,\, J_{\mathfrak{m}_x} \mkern-1mu=\mkern-1mu K_{\mathfrak{m}_x} \,\,\, \text{for all maximal ideals } \mathfrak{m}_x,
  \label{eq:LocalEquality}
\end{equation}
where $J_{\mathfrak{m}_x}$ is the image of $J$ in the localization of the polynomial
ring at $\mathfrak{m}_x$ and we make use of the fact that, in an algebraically
closed field, maximal ideals can be labeled by points $x$ in affine space.
The local nature of ideal containment can be easily argued by applying
\cite[Proposition 3.8]{AtiyahMacdonald1969} to the module $M = (J+K)/J$.
Intuitively, \cref{eq:LocalEquality} tells us to check that
\cref{eq:A0Decomposition} holds locally at all points.

The localization strategy is fruitful as the actors in \cref{eq:A0Decomposition}
behave nicely under localization. Firstly, one has 
\begin{equation}
J^{[U]}_{\mathfrak{m}_x} = J_{\mathfrak{m}_x}, \quad \text{for all } x \in U.
\end{equation}
As such, $J^{[U]}$ is the pre-image of the localization of $J$ at any point $x
\in U$, hence the name.
Secondly, Fitting ideals commute with localization, i.e., one has that
\begin{equation}
  \text{Fitt}_0(M)_{\mathfrak{m}_x} = \text{Fitt}_0(M_{\mathfrak{m}_x}).
\end{equation}
This allows us to make frequent use of 
cases where Fitting ideals can easily be understood. Specifically, for an
ideal $J$ one has that
\begin{align}
  \text{Fitt}_0(\langle 0 \rangle_{\mathfrak{m}_x}) &= R_{\mathfrak{m}_x},
  \label{eq:FittingIsRing}
  \\
   \text{Fitt}_0(R_{\mathfrak{m}_x}/J_{\mathfrak{m}_x}) &= J_{\mathfrak{m}_x}.
  \label{eq:FittingIsIdeal}
\end{align}
This collection of properties, together with the fact that
\cref{eq:A0Decomposition} is organized by the Landau locus are all that is
required for the proof.
Specifically, we break down the application of \cref{eq:LocalEquality} into
three cases:
\begin{enumerate}[itemsep=0mm]
  \item $x \not\in U_{\text{crit}[\log(B)]}^\Gamma \cup U_{\text{Landau}}^\Gamma$,
  \item $x \in U_{\text{crit}[\log(B)]}^\Gamma$ and
  \item $x \in U_{\text{Landau}}^\Gamma$,
\end{enumerate}
where $U_{\text{crit}[\log(B)]}^\Gamma$ is the
critical locus of the logarithm of the maximal-cut Baikov polynomial.

First, we argue that \cref{eq:A0Decomposition} holds locally on
$x \not\in U_{\text{crit}[\log(B)]}^\Gamma \cup U_{\text{Landau}}^\Gamma$.
To this end, note that
    \begin{equation}
        U_{\text{crit}[\log(B)]}^\Gamma \cup U_{\text{Landau}}^\Gamma = U_{\text{syz}}^\Gamma \cap U_{\text{cut}}^\Gamma.
    \end{equation}
If we consider localizing at a point outside of this variety then either
$x \not\in U_{\text{cut}}^\Gamma$ and
$(J_{\text{cut}}^\Gamma)_{\mathfrak{m}_x} = R_{\mathfrak{m}_x}$ or $x \not\in
U_{\text{syz}}^\Gamma$ and $\text{Fitt}_0(M_i)_{\mathfrak{m}_x} =
\text{Fitt}_0(\langle 0 \rangle_{\mathfrak{m}_x}) = R_{\mathfrak{m}_x}$.
In each case, both sides of \cref{eq:A0Decomposition} are locally
$R_{\mathfrak{m}_x}$ and equality holds.

Second, we consider localizing at $x \in U_{\text{crit}[\log(B)]}^\Gamma$.
As $B \ne 0$ on $U_{\text{crit}[\log(B)]}^\Gamma$, and $B = 0$ on both $V(J^{\Gamma,[U_i]}_{\text{syz}})$
and $V(J_{\text{4d}})$, localizing at $x$ gives
\begin{equation}
  \text{Fitt}_0(M_i)_{\mathfrak{m}_x}= \text{Fitt}_0(R_{\mathfrak{m}_x}/[J_{\text{syz}}^\Gamma]_{\mathfrak{m}_x}) =
  (J_{\text{syz}}^\Gamma)_{\mathfrak{m}_x}.
\end{equation} Moreover, as $B \ne 0$ on $x$, $B$ is a unit
in $R_{\mathfrak{m}_x}$ and so
$(J_{\text{syz}}^\Gamma)_{\mathfrak{m}_x} = (A_0^\Gamma)_{\mathfrak{m}_x}$.
Therefore, \cref{eq:A0Decomposition} holds locally here.

Third, we show local equality for $x \in U_{\text{Landau}}^\Gamma$.
For the left-hand side, note that $B=0$ on $x$, and so we find
$R_{\mathfrak{m}_x}$.
For the right-hand side, we consider the decomposition of
$U_{\text{Landau}}^\Gamma$ in \cref{eq:LandauDecomposition}. First, we consider
the localization at a point on some $U_i$. By construction,
$(M_i)_{\mathfrak{m}_x} = \langle 0 \rangle_{\mathfrak{m}_x}$ and so
$\text{Fitt}_0(M_i)_{\mathfrak{m}_x} = R_{\mathfrak{m}_x}$.
Next, consider the localization on any point of $U_0^\Gamma$ that is not
contained within the $U_i$.
By \cref{eq:RadicalityCondition}, $(J_{\text{syz}}^\Gamma)_{\mathfrak{m}_x}$ is
radical on these points, and hence equal to $(J_{\text{4d}})_{\mathfrak{m}_x}$. Locally, we see that each $(M_i)_{\mathfrak{m}_x} = \langle 0
\rangle_{\mathfrak{m}_x}$ and so each $\text{Fitt}_0(M_i) =
R_{\mathfrak{m}_x}$, and so \cref{eq:A0Decomposition} is proven.
Finally, as this last statement is true as long as there is at least one term in
the sum over $i$ in \cref{eq:A0Decomposition}, we can drop the $i=0$ term if
$n_{\text{IR}} > 0$.

\section{Remaining $pp \rightarrow t\overline{t}H$ Five-Point Examples}
In this appendix, we provide ingredients for the Landau decompositions for the
five-point integrals in \cref{fig:ExtraExamples}.

\paragraph{Pentabox}
We consider $\Gamma_2$, the pentabox of \cref{fig:PB}.
Here, the maximal cut Landau locus is
\begin{equation}
  U^{\Gamma_2}_{\text{Landau}} = U_{0}^{\Gamma_2} \cup U_{1}^{\Gamma_2} \cup U_{2}^{\Gamma_2}.
\end{equation}
On $U_{1}^{\Gamma_2}$, the momentum in the line between $p_4$ and $p_5$ is
soft and on $U_{2}^{\Gamma_2}$, the momentum in central line is soft. In both
cases, the remaining propagators are on-shell.

If we pre-localize at $U_{1}^{\Gamma_2}$ and intersect with $J_{\text{4d}}$, we find the ideal in
\cref{eq:ExternalSoft} and no new computation is required.
In contrast, the pre-localization at $U_{2}^{\Gamma_2}$ is novel.
We find
\begin{align}
  \begin{split}
    \!\!\! J_{\text{syz}}^{\Gamma_2, [U_{2}^{\Gamma_2}]} \!\cap\! J_{\text{4d}} \!=\! \langle \Lambda^\nu_{1 \perp_{\!P}}\!, z_{1,0} (\overline{\Lambda}_1 \! \cdot \! P), z_{1,3} (\Lambda_1 \! \cdot \! P) &, \tilde{z} \mu_{11}, \\
 \tilde{z} \mu_{12}, \mu_{11} \mu_{22} - \mu_{12}^2 \rangle + (1 &\leftrightarrow 2).
  \end{split}
  \label{eq:PentaboxPrelocalization}
\end{align}
where $P = q_3$, $\overline{\Lambda}_k = \Lambda_k|_{\ell_i \rightarrow P - \ell_i}$ and $v_{\mu}^{\perp_{\!P}} = v_{\mu}
- \frac{(v \cdot P)}{P^2}P_\mu$ is the projection of $v$ onto the space
transverse to $P$.
The $\mathcal{C}$-matrix can be constructed from \cref{eq:dZmu12} as well
as
\begin{equation}
  \partial_{i,j} B = w_j \cdot \Lambda_i^{\perp_{\!P}}, \qquad \text{for} \quad j \in \{1, 2, 4\},
\end{equation}
and the four relations generated by applying $1 \leftrightarrow 2$ and $\ell_i
\rightarrow P - \ell_i$ to
\begin{equation}
  z_{1,3} \partial_{1,3} B = z_{1,3} (w_3 \cdot \Lambda_1^{\perp_{\!P}}) + \frac{(w_3 \cdot P)}{P^2} z_{1,3} (\Lambda_1 \cdot P).
\end{equation}

\paragraph{Pentatriangle}
We consider the pentatriangle diagram of \cref{fig:PT}. The Landau locus
decomposes as
\begin{equation}
  U_{\text{Landau}}^{\Gamma_3} = U_{0}^{\Gamma_3} \cup U_1^{\Gamma_3} \cup U_2^{\Gamma_3} \cup U_3^{\Gamma_3}.
\end{equation}
On $U_1^{\Gamma_3}$, the central line momentum is soft; on $U_2^{\Gamma_3}$,
$\ell_2$ is soft; on $U_3^{\Gamma_3}$, $\ell_2$ is collinear to $p_5$. In all
cases, the other line momenta are on-shell.
In practice, it turns out that the Fitting ideal constructed from
$U_1^{\Gamma_3}$ contains the other possible Fitting ideals, so we focus here,
finding
\begin{equation}
  J_{\text{syz}}^{[U_1^{\Gamma_3}]} \cap J_{\text{4d}} = \langle \Lambda_1^\nu, \Lambda_2 \cdot p_5, \Lambda_2^{\perp \nu}, B, \tilde{z} \mu_{22}, \chi_i^{(a)} \rangle.
  \label{eq:DoubleCollinearPrelocalization}
\end{equation}
To construct the $\mathcal{C}$ matrix, we first note that the Baikov polynomial
is manifestly in the ideal. Further rows can be constructed by using
\cref{eq:dzi0LambdaMaRelation} for $i = 1$ and
\cref{eq:BaikovDerivativesMomenta} for $a = 1$. Next we note that for $j = \{1,2,3\}$
\begin{equation}
  \partial_{2,j} B = w_j \cdot \Lambda_2.
\end{equation}
Moreover, one has that
\begin{equation}
  \tilde{z}\partial_{\tilde{z}}B \!=\! \ell_1 \cdot \Lambda_1^\perp + 2 \frac{(\ell_1 \!\cdot\! \eta)(\Lambda_1 \!\cdot\! p_5)}{p_5 \!\cdot\! \eta} + B - \chi_1^{(a)} + (1 \!\leftrightarrow\! 2).
\end{equation}
For the remaining $ z_{2,j} \partial_{2,j}B$ we start by rewriting
\begin{align}
  \begin{split}
    z_{2,j} \partial_{2,j} B &=
\frac{z_{2,j}}{\eta \!\cdot\! p_5}\left[ (w_j \!\cdot\! \eta) (p_5 \!\cdot\! \Lambda_2) + (w_j \!\cdot\! p_5) (\eta \!\cdot\! \Lambda_2) \right]
     \\
  &\quad + \chi_2^{(a)} \delta_{j,0} + z_{2,j} (w_j \!\cdot\! \Lambda_2^\perp) ,
  \end{split}
\end{align}
where only the last term is not manifestly in the ideal of
\cref{eq:DoubleCollinearPrelocalization}. To remedy this, we first define
\begin{equation}
  u_j^\nu = \frac{2(\eta \cdot [\ell_2 - q_j])}{\eta \cdot p_5} p_5^\nu + (\ell_2^{\perp})^\nu.
\end{equation}
This allows us to write
\begin{equation}
  z_{2,j} (\eta \cdot \Lambda_{2}) = (u_j \cdot \ell_2) (\eta \cdot \Lambda_2) + \mu_{22} (\eta \cdot \Lambda_2),
  \label{eq:DoubleCollinearDetailedRewriting}
\end{equation}
which holds for $j \in \{0,4\}$.
The right-hand side of \cref{eq:DoubleCollinearDetailedRewriting} can be written in
terms of the generators in \cref{eq:DoubleCollinearPrelocalization} via
\begin{equation}
\mu_{22} (\Lambda_2 \cdot \eta) = B (\ell_2 \cdot \eta) - \mu_{12} (\Lambda_1 \cdot \eta),
\end{equation}
\vspace{-7mm}
\begin{equation}
  [u \cdot \ell_2] [\eta \cdot \Lambda_2] \!=\! [u \cdot \Lambda_2] [\eta \cdot \ell_2] \mkern-1mu-\mkern-1mu [u \cdot \ell_1] [\eta \cdot \Lambda_1] \mkern-1mu+\mkern-1mu [\eta \cdot \ell_1] [u \cdot \Lambda_1].
  \end{equation}

\paragraph{First Box-Triangle} We consider diagram $\Gamma_4$ of \cref{fig:BT}.
The maximal-cut Landau locus decomposes as
\begin{equation}
  U_{\text{Landau}}^{\Gamma_4} = U_{0}^{\Gamma_4} \cup U_{1}^{\Gamma_4} \cup U_{2}^{\Gamma_4}.
\end{equation}
On $U_{1}^{\Gamma_4}$, $\ell_2$ is collinear to $p_5$ and on $U_{2}^{\Gamma_4}$
the middle line momentum is soft. In both cases, the remaining propagators of
the diagram are on-shell.

To pre-localize at $U_{1}^{\Gamma_4}$, we make use of the Sudakov
parameterization for $p = p_5$, which leads to
\begin{equation}
    J_{\text{syz}}^{\Gamma_4, [U_{1}^{\Gamma_4}]} \cap J_{\text{4d}} \!=\! \langle \mu_{12}, \mu_{22}, \mu_{11}  \ell_2^{\perp \nu},   z_{2,0} \mu_{11}, z_{2,4} \mu_{11} \rangle.
  \label{eq:ExternalColIdeal}
\end{equation}
To construct $\mathcal{C}$, we express the generators of
$J_{\text{syz}}^{\Gamma_4}$ in terms of those in \cref{eq:ExternalColIdeal}
using \cref{eq:dZmu12} and
\begin{align}
  &\partial_{1,j} B = (\partial_{1,j} \mu_{11}) \mu_{22} - 2 (\partial_{1,j} \mu_{12}) \mu_{12},
    \label{eq:SimpledZ1Derivatives}
  \\
  &z_{2,j} \partial_{2,j} B = (\partial_{2,j} \mu_{22}) z_{2,j} \mu_{11} - 2 z_{2,j} (\partial_{2,j} \mu_{12}) \mu_{12},
  \label{eq:CollinearMembership1}
  \\
  &\partial_{2, j} B = \left[ \! \frac{\eta  \!\cdot\! w_j}{2\eta \!\cdot\! p_5}(z_{2,4} \!-\! z_{2,0}) + w_j \!\cdot\! \ell_2^\perp \! \right] \! \mu_{11} \!-\! \partial_{2,j} \mu_{12}^2,
  \label{eq:CollinearMembership2}
\end{align}
where \cref{eq:CollinearMembership2} holds for $j \in \{1, 2, 3\}$.
Pre-localizing at $U_{2}^{\Gamma_4}$ gives
\begin{equation}
  J_{\text{syz}}^{\Gamma_4, [U_{2}^{\Gamma_4}]} \cap J_{\text{4d}} = \langle \mu_{11} - \mu_{12}, \mu_{11} - \mu_{22}, \mu_{11}(\ell_1 - \ell_2)^\nu \rangle
  \label{eq:InternalOnlySoft}
\end{equation}
and the associated $\mathcal{C}$-matrix can be constructed by noting 
\begin{align}
  \begin{split}
    \tilde{z} \partial_{\tilde{z}} B &= (\mu_{11} \!+\! \mu_{22} \!-\!\tilde{z})(\mu_{11} \!-\! \mu_{12}) - \mu_{12}(\mu_{11} \!-\! \mu_{22}) \\
                                       &\quad + (\ell_{1} \!-\! \ell_2) \cdot [\mu_{11} (\ell_1 \!-\! \ell_2)],
  \end{split}
    \\
  \Lambda_{1}^\nu &= \mu_{11}(\ell_1 \!-\! \ell_2)^\nu + \ell_2^\nu (\mu_{11} \!-\! \mu_{12}) - \ell_1^\nu(\mu_{11} \!-\! \mu_{22}), \\
  \Lambda_{2}^\nu &= \ell_1^\nu (\mu_{11} \!-\! \mu_{12}) - \mu_{11}(\ell_1 \!-\! \ell_2)^\nu,
\end{align}
from which it is easy to construct the rows of the $\mathcal{C}$-matrix by
comparison with \cref{eq:BaikovDerivativesMomenta}.

\paragraph{Second Box-Triangle}
We consider diagram $\Gamma_5$ of \cref{fig:BT2}.
The maximal-cut Landau locus decomposes as
\begin{equation}
  U_{\text{Landau}}^{\Gamma_5} = U_{0}^{\Gamma_5} \cup U_{1}^{\Gamma_5} \cup U_{2}^{\Gamma_5},
\end{equation}
where on $U_{1}^{\Gamma_5}$ the central line momentum is soft and on
$U_{2}^{\Gamma_5}$ the line momentum exchanged between $p_4$ and $p_5$ is soft.
All other edges are on-shell. If we pre-localize at $U_{1}^{\Gamma_5}$ and
intersect with $J_{\text{4d}}$, we find the ideal in \cref{eq:InternalOnlySoft}.
If we pre-localize at $U_{2}^{\Gamma_5}$ and
intersect with $J_{\text{4d}}$, we find the ideal in \cref{eq:ExternalSoft}.
As such, in both cases, no further computation is required.

\paragraph{Pentabubble}
We consider the pentabubble diagram of \cref{fig:PBub}.
The maximal-cut Landau locus decomposes as
\begin{equation}
  U_{\text{Landau}}^{\Gamma_6} = U_{0}^{\Gamma_6} \cup U_1^{\Gamma_6} \cup U_2^{\Gamma_6}.
\end{equation}
On $U_1^{\Gamma_6}$, the central line momentum is soft, and the remaining lines
are on-shell.
The pre-localization on $U_1^{\Gamma_6}$ is the ideal given in
\cref{eq:InternalOnlySoft}. The $\mathcal{C}$ matrix can then be built from the
corresponding discussion.

On $U_2^{\Gamma_6}$ the line momentum associated to $z_{2,4}$ is soft with the
remaining lines on-shell. Here, the pre-localization computation is novel. One
finds
\begin{equation}
    J_{\text{syz}}^{\Gamma_6, [U_2^{\Gamma_6}]} \cap J_{\text{4d}} = \langle \mu_{22}, \mu_{12}, \Lambda_2^\nu  + (\mu_{11} - \mu_{12}) p_5^\nu \rangle.
    \label{eq:PentabubbleExternalSoft}
\end{equation}
The relations required to express the $\mathcal{C}$ matrix are
\cref{eq:dZmu12,eq:SimpledZ1Derivatives} as well as
\begin{align}
  \begin{split}
    z_{2,4} \partial_{2,4} B &= ([\ell_1 \!-\! \ell_2] \!\cdot\! [\ell_2 \!+\! p_5] \!-\! \mu_{22}) \mu_{12} \!+\! \mu_{11} \mu_{22}\\
                               & \quad  \!+ \!(\ell_2 \!+\! p_5 \!+\! z_{2,4} w_4) \!\cdot\! (\Lambda_2 \!+\! [\mu_{11} \!-\! \mu_{12}] p_5),
  \end{split}
  \\
  \partial_{2,i} B &= w_i \cdot (\Lambda_2 + p_5[\mu_{11} - \mu_{12}]),
\end{align}
where the second relation holds for $i \ne 4$.

\section{Relation Matrix Ancillary Files}

Expressions for the $\mathcal{R}_i$ submatrices needed to construct each of the
$\mathcal{F}_i$ studied in this work can be found in Mathematica format under
\texttt{relationMatrices/*.m}.
Each file is a pair of generators of a pre-localization ideal and syzygies of
the generators. There is one file for each of the 7 distinct pre-localizations
that we discuss, organized as
\begin{itemize}
  \item \texttt{soft.m}, corresponding to \cref{eq:PentabubbleExternalSoft};
  \item \texttt{col.m}, corresponding to \cref{eq:ExternalColIdeal}, for \texttt{p}
    $= p_5$;
  \item \texttt{2Col.m}, corresponding to \cref{eq:ExternalSoft} for \texttt{(p,eta)}
    $= (p_5, p_4)$;
  \item \texttt{midSoft.m}, corresponding to \cref{eq:InternalOnlySoft};
  \item \texttt{midSoftCol.m}, corresponding to
    \cref{eq:InternalSoftCol};
  \item \texttt{midSoft2Col.m}, corresponding to \cref{eq:PentaboxPrelocalization};
  \item \texttt{pt.m}, corresponding to
    \cref{eq:DoubleCollinearPrelocalization}, for \texttt{p} 
    $= p_5$ and using a Sudakov parameterization everywhere.
\end{itemize}
In the files, we represent the loop momentum components $\ell_i^\nu$ as
\texttt{li[nu]}, $\mu_{ij}$ as \texttt{muij} and scalar products of
vectors $v \cdot w$ as \texttt{sp[v, w]}.
For the four-dimensional momentum space, we use the alternating metric
signature, $+-+-$.
We represent the propagator variables $z_{a,j}$ as 
\texttt{z[a,j]} and $\tilde{z}
$ as \texttt{zt}.
To represent the Sudakov parameterization, we denote $x_i$ as
\texttt{xi} and $\beta_i$ as \texttt{bei}. Moreover, we explicitly pick a
basis of the transverse space, writing
\begin{equation}
  \ell_i^\perp = \alpha_{i,+} \omega_{+} + \alpha_{i,-} \omega_{-},
\end{equation}
where we make use of $\omega_{\pm}$ such that
\begin{equation}
  \omega_{\pm}^2 = \omega_{\pm} \cdot p = \omega_{\pm} \cdot \eta = 0, \qquad \omega_{+} \cdot \omega_{-} = - \eta \cdot p.
\end{equation}
In the files, we represent $\alpha_{i,+}$ as \texttt{alip}
and $\alpha_{i,-}$ as \texttt{alim}.
For the pentabox example, we employ a basis for the space transverse to $P$, we
introduce three four-vectors $n_i$ that satisfy
\begin{equation}
  n_i\cdot P = 0, \quad n_i \cdot n_j = \delta_{ij},
\end{equation}
where we represent the $n_i$ as \texttt{n[i]}. For simplicity, we abbreviate
$(\overline{\ell}_a - P)^2$ as \texttt{z[a,P]}.

\bibliography{criticalsyz2}

\end{document}